# A Decidable Probability Logic for Timed Probabilistic Systems

Ruggero Lanotte[a] and Danièle Beauquier[b]

[a]*Dipartimento di Scienze della Cultura, Politiche e dell'Informazione, Università dell'Insubria*

[b]*LACL, Dept. of Informatics Universitè de Paris 12*

**Abstract**

In this paper we extend the predicate logic introduced in [Beauquier et al. 2002] in order to deal with Semi-Markov Processes. We prove that with respect to qualitative probabilistic properties, model checking is decidable for this logic applied to Semi-Markov Processes. Furthermore we apply our logic to Probabilistic Timed Automata considering classical and urgent semantics, and considering also predicates on clocks. We prove that results on Semi Markov Processes hold also for Probabilistic Timed Automata for both the two semantics considered. Moreover, we prove that results for Markov Processes shown in [Beauquier et al. 2002] are extendable to Probabilistic Timed Automata where urgent semantics is considered.

## 1 Introduction

This work is in keeping with the general pattern of specification and verification of real time systems. Among the numerous existing frameworks within which a formal analysis can be carried out, the *timed automata* formalism [1] has received much attention.

Classically the timed properties to verify are expressed in terms of temporal logics. Moreover during the last years, a new feature has become of interest, namely the possible probabilistic behavior of a real system (there is a large field of application to fault tolerant systems, timed randomized algorithms and in communication protocols, see [2], [3] and [4]). As a result, several models of probabilistic timed automata have been developed (see [5] and [6]) as well as a lot of probabilistic temporal logics, and automatic verification methods for these models against the respective logics.

* *Corresponding author*: Ruggero Lanotte: Dipartimento di Scienze della Cultura, Politiche e dell'Informazione, Università dell'Insubria, Via Valleggio 11, 22100, Como, Italy, ruggero.lanotte@uninsubria.it.
Research partially supported by MIUR Progetto Cofinanziato "Modelli Formali per la Sicurezza e il Tempo" (MEFISTO).

A timed Automata is a finite state machine equipped with real variables called clocks. A transition is instantaneous and is triggered by a set of values for the clocks expressed by formulae of the form $x \sim c$ and $x - y \sim c$. When a transition is taken it can reset to 0 the value of a certain set of clocks. The values of clocks are increased with the time elapsed in the state before performing a transition. A probabilistic Timed Automata is a Timed automata where a discrete probability is associated to the choice of the transition that can be taken from a state. In the literature (see among the others [7], [8], [9], [10] and [11]) also urgent semantics are considered. Urgency is necessary to model deadlines and systems that must react as soon as possible to a certain stimuli.

Recently, in [12] a predicate logic of probabilities has been studied which leads to decidable model checking, when applied to Finite Probabilistic Processes (i.e. finite labelled Markov chains [13]). Finite Probabilistic Processes do not involve non determinism, contrary to Semi-Markov Processes which include both non determinism and probabilities. This model naturally implies the notion of adversary (or policy, strategy, depending on the authors). The adversary is used to resolve the non determinism.

Qualitative probabilistic properties (probabilities are 0 or 1) are properties that are fulfilled by almost all executions, and hence are largely studied (see among the others [14] and [15]) since allow to express liveness/fariness properties for probabilistic systems.

In this paper we extend the predicate logic introduced in [12] in order to deal with Semi-Markov Processes, by a modification of the probabilistic operators. We prove that with respect to qualitative probabilistic properties, model checking is decidable for this logic applied to Semi-Markov Processes.

Furthermore we apply our logic to Probabilistic Timed Automata (considering classical and urgent semantic) giving to some predicates a fixed semantics: these predicates are *clock predicates*. They are of the form $x_{t_1} - y_{t_2} + c$ where $z_t$ is the real value of clock $z$ at step $t$ ($t$ is a natural), and $c$ is an integer. We obtain two kinds of results: undecidability and decidability ones. In the general case, even without probabilistic operators, it turns out that model checking is undecidable. If one restricts to clock predicates of the form $x_t - y_t + c$, then, firstly, qualitative model checking is decidable, secondly, for urgent semantics, quantitative model checking is decidable for "almost" all values of the probabilistic parameters.

The structure of the paper is as follows. Section 2 gives basic definitions about labelled transition systems and weak second order monadic logic of order. In Section 3 we define a new logic and prove that model checking is decidable for this logic applied to Semi-Markov Processes. Section 4 is devoted to undecidability and decidability results concerning model checking for Probabilistic Timed Automata with this logic enriched with clock predicates. The last section describes the future work and compares our logic with the existing ones.

## 2  Basic notions

A *labelled transition system* $S$ is a tuple $(\Delta, L, Q, q_0, Tr, \lambda)$ such that: $\Delta$ is a set of symbols, $L$ is a set of atomic propositions, $Q$ is a set of states, $q_0 \in Q$ is the initial state, $Tr \subseteq Q \times \Delta \times Q$ is a set of *transitions*, and $\lambda$ is a function assigning to each state $q$ a subset of atomic propositions $\lambda(q)$. If $q$ is a state and $a$ is a symbol, then with $S(q, a)$ we denote the set of transitions with source $q$ and symbol $a$ of the labelled transition system $S$, more precisely it is the set $\{(q, a, q') \mid (q, a, q') \in Tr\}$. We require that the set $S(q, a)$ is finite; for each state $q$ and symbol $a$. Hence the set of states $Q$ and the set of symbols $\Delta$ can be infinite, but the set of transitions $S(q, a)$ must be finite.

A *run* of $S$ is a possible infinite sequence of steps of the form $\omega = q_1 \xrightarrow{a_1} q_1 \xrightarrow{a_2} \ldots$ where $(q_i, a, q_{i+1})$ is in $Tr$. The length of $\omega$, denoted $length(\omega)$, is equal to $n$ if $\omega$ is the finite run $q_1 \xrightarrow{a_1} \ldots q_n \xrightarrow{a_n} q_{n+1}$, and $\infty$ otherwise.

If $length(\omega) = \infty$, then with $inf(\omega)$ we denote the set of states of $S$ crossed in $\omega$ infinitely many times. Moreover, let $k \leq length(\omega)$; with $\omega(k)$ we denote the state $q_k$ and with $\omega^{(k)}$ we denote the run $q_1$ if $k = 0$, and the run $q_1 \xrightarrow{a_1} \ldots q_k \xrightarrow{a_k} q_{k+1}$, otherwise.

If $k = length(\omega)$, then we say that $\omega$ *is a prefix of* $\omega'$ if and only if $length(\omega') \geq k$ and $\omega = (\omega')^{(k)}$.

If $\omega$ is a run $q_1 \xrightarrow{a_1} \ldots q_{n-1} \xrightarrow{a_{n-1}} q_n$, $q$ is a state and $a$ is a symbol, sometimes we will write $\omega \xrightarrow{a} q$ to denote the run $q_1 \xrightarrow{a_1} \ldots q_{n-1} \xrightarrow{a_{n-1}} q_n \xrightarrow{a} q$.

With $Path_{fin}(S, q)$ (resp. $Path_{ful}(S, q)$) we denote the set of finite (resp. infinite) runs $\omega = q_1 \xrightarrow{a_1} \ldots \xrightarrow{a_n} q_{n+1} \ldots$ of $S$ such that $q = q_1$. Moreover, with $Path_{fin}(S)$ and $Path_{ful}(S)$ we denote the sets $Path_{fin}(S, q_0)$ and $Path_{ful}(S, q_0)$, namely the set of finite and infinite runs starting from the initial state $q_0$.

A labelled transition system $S = (\Delta, L, Q, q_0, Tr, \lambda)$ is a *Finite Automaton* if $\Delta$, $L$ and $Q$ are finite.

We recall the syntax and semantics of the Weak Monadic Logic of Order ($WMLO$)

**Definition 1** *Let $Z$ be a set of monadic predicate symbols; the set $WMLO(Z)$ is the set of formulae $\phi$ on $Z$ defined by the following grammar:*

$$\phi ::= B(t) \mid t < t' \mid t \in X \mid \exists t.\phi_1 \mid \exists X.\phi_1 \mid \neg \phi_1 \mid \phi_1 \vee \phi_2$$

*where $B$ is a monadic predicate symbol in $Z$, $t, t'$ are two natural variables, $X$ is a variable representing a finite set of naturals.*
*Conjunction, implication and universal quantification can be easily derived.*

We give now the semantics of a $WMLO(Z)$-formula over the structure $(\mathbb{N}, <)$.

A *valuation* $v$ is a function that assigns to each predicate symbol $B \in Z$ a subset of $\mathbb{N}$, to each variable $t$ a natural, and to each variable $X$ a *finite* subset of $\mathbb{N}$. With $v[n/t]$ and $v[N/X]$ we denote the valuation that coincides with $v$ except it assigns the value $n$ to variable $t$ and the value $N$ to variable $X$ respectively.

**Definition 2** *We define when a formula $\phi$ holds on $(\mathbb{N}, <)$ under a valuation $v$, written $\mathbb{N}, v \models \phi$, by the following inductive clauses:*

$\mathbb{N}, v \models B(t)$      *iff*   $v(t) \in v(B)$

$\mathbb{N}, v \models t < t'$      *iff*   $v(t) < v(t')$

$\mathbb{N}, v \models t \in X$      *iff*   $v(t) \in v(X)$

$\mathbb{N}, v \models \exists t.\phi_1$      *iff*   $\mathbb{N}, v[n/t] \models \phi_1$, *for some $n \in \mathbb{N}$*

$\mathbb{N}, v \models \exists X.\phi_1$      *iff*   $\mathbb{N}, v[N/X] \models \phi_1$, *for some finite set of naturals $N$*

$\mathbb{N}, v \models \neg \phi_1$      *iff*   *it is not the case $\mathbb{N}, v \models \phi_1$*

$\mathbb{N}, v \models \phi_1 \vee \phi_2$      *iff*   *either $\mathbb{N}, v \models \phi_1$ or $\mathbb{N}, v \models \phi_2$*

We note that, we can express constants $0, 1, \ldots$ and conditions of the form $t \sim c$ and $t \sim t' + c$, where $t$ and $t'$ are natural variables, $c$ is a natural and $\sim \in \{<, \leq, =, \neq, \geq, >\}$. For more details see [16] and [17].
This is a classical fact that finite automata have the same expressive power as the $WMLO$ logic. We make it more precise below.
If $S$ is a Finite Automaton and $F$ is a subset of states of $S$, then with $\mathcal{L}(S, F)$, we denote the set of infinite words $a_0 a_1 \ldots$ such that there exists a infinite run $\omega = q_0 \xrightarrow{a_0} q_1 \xrightarrow{a_1} \ldots$ with $inf(\omega) \cap F \neq \emptyset$ (this is the Büchi acceptance condition; see [17]).

Let $L = \{B_1, \ldots, B_k\}$ be a set of atomic propositions, and $\phi$ be a $WMLO(L)$-formula with free variables in $\mathcal{X} = \{t_1, \ldots, t_n, X_1, \ldots, X_m\}$. The values assigned by a valuation $v$ to $L$ and $\mathcal{X}$ can be represented by the infinite word $\alpha(v) = a_0 a_1 \ldots$ on the alphabet $\Delta = \{0,1\}^{k+n+m}$ in the following way: if $a_j = (b_1^j, \ldots, b_{k+n+m}^j)$ and

$$(S_1, S_2, \ldots, S_{k+n+m}) = (B_1, \ldots, B_k, \{v(t_1)\}, \ldots, \{v(t_n)\}, v(X_1), \ldots, v(X_m)),$$

then $j \in v(S_i)$ iff $b_i^j = 1$.

With $\mathcal{L}(\phi)$ we denote the set of infinite words $\alpha(v)$ such $\mathbb{N}, v \models \phi$.

The following theorem can be derived from the results given in [16] and [17].

**Theorem 3** *Let $L = \{B_1, \ldots, B_k\}$ be a set of atomic propositions and $\phi \in WMLO(L)$ with free variables in $\mathcal{X} = \{t_1, \ldots, t_n, X_1, \ldots, X_m\}$, one can compute a Finite Automaton $S$ and a subset of states $F$ such that $\mathcal{L}(S, F) = \mathcal{L}(\phi)$, and vice versa.*

## 3 Probabilistic extensions

In this section, we consider *Probabilistic Structures* and a *Probabilistic Monadic Logic of Order PMLO* for them. We recall the definition of *Semi Markov Processes* as a Probabilistic Structure, we recall known results about decidability for *Markov Process* (a sub class of the class of Semi Markov Processes). Moreover, we prove a decidability result for the class of Semi Markov Processes.

### 3.1 Probabilistic Structures

**Definition 4** *A Probabilistic Structure is a pair $M = (S^M, \rho^M)$, where $S^M = (\Delta, L, Q, q_0, Tr, \lambda)$ is a labelled transition system and $\rho^M : Tr \to (0,1]$ is a probability transition function such that for each state $q \in Q$ and each symbol $a \in \Delta$ we have that $\sum_{e \in S^M(q,a)} \rho^M(e) = 1$. We note that a transition cannot have a probability equal to 0.*

From now on, for simplicity, we do not make distinction between $M$ and the labelled transition system $S^M$ of $M$. As an example we will write $Path_{fin}(M)$ to denote the set $Path_{fin}(S^M)$, and $M(q,a)$ to denote $S^M(q,a)$.

**Definition 5** *An adversary $A$ of a Probabilistic Structure $M$ is a function from $Path_{fin}(M)$ to $\Delta$ such that if $A(\omega) = a$, then there exists a state $q'$ such that $\omega \xrightarrow{a} q'$ is in $Path_{fin}(M)$.*

If $A$ is an adversary of $M$, then with $Path^A_{fin}(M)$ (resp. $Path^A_{ful}(M)$) we denote the set of finite (resp. infinite) runs $\omega = q_0 \xrightarrow{a_0} \ldots \xrightarrow{a_n} q_{n+1} \ldots$ of $M$ such that $a_i = A(\omega^{(i)})$, for any $0 \leq i < length(\omega)$.

**Definition 6** *If $\omega$ is a finite run $q_0 \xrightarrow{a_0} \ldots q_{n-1} \xrightarrow{a_{n-1}} q_n$, then with $\overline{\mu}(\omega)$ we denote the probability computed as follows*

$$\overline{\mu}(\omega) = \begin{cases} 1 & \text{if } n = 0 \\ \overline{\mu}(\omega^{(n-1)}) \cdot \rho^M((q_{n-1}, a_{n-1}, q_n)) & \text{if } n > 0 \end{cases}.$$

If $A$ is an adversary of $M$, then with $F^A_{path}(M)$ we denote the smallest $\sigma-$algebra on $Path^A_{ful}(M)$ that contains the sets

$$\{\omega \mid \omega \in Path^A_{ful}(M) \ \land \ \omega' \text{ is a prefix of } \omega\}$$

for any $\omega' \in Path^A_{fin}(M)$.

**Definition 7** *The measure $\mu$ on the $\sigma-$algebra $F^A_{path}(M)$ is the unique measure such that*

$$\mu(\{\omega \mid \omega \in Path^A_{ful}(M) \ \land \ \omega' \text{ is a prefix of } \omega\}) = \overline{\mu}(\omega')$$

for any $\omega' \in Path^A_{fin}(M)$.

**Definition 8** *Let $Z$ be a set of monadic probabilistic predicate symbols; the set $PMLO(Z)$ is the set of formulae $\phi$ on $Z$ defined by the following grammar:*

$$\phi ::= B(t) \mid t < t' \mid t \in X \mid \exists P_{\sim p}(\phi_1 | \phi_2) \mid \exists t.\phi_1 \mid \exists X.\phi_1 \mid \neg \phi_1 \mid \phi_1 \lor \phi_2$$

*where $B$ is a monadic probabilistic predicate symbol in $Z$, $t, t'$ are two natural variables, $X$ is a variable representing a finite set of naturals, $\sim \in \{<, \leq, =, \neq, \geq, >\}$ and $p$ is a rational number in $[0, 1]$.*
*Conjunction, implication and universal quantification can be easily derived. We call $\exists P_{\sim p}(\_)$ the probabilistic operator. With $\exists P_{\sim p}(\phi)$ we will denote the formula $\exists P_{\sim p}(\phi | true)$.*

The probabilistic operator $\exists P_{\sim p}(\phi_1 | \phi_2)$ means that there exists an adversary of $M$ such that the probability that $\phi_1$ holds when $\phi_2$ holds is related with the rational $p$ with the relation $\sim \in \{<, \leq, =, \neq, \geq, >\}$. A formula in $PMLO(Z)$ is *closed* if and only if it has no free variables.

We give now the semantics of a $PMLO(Z)$-formula over a probabilistic structure $M$ where $Z$ is its set of atomic propositions. These atomic propositions are considered as probabilistic monadic predicates, and a valuation $v$ for them assigns to each run $\omega$ of $M$ a subset of $\mathbb{N}$ in the following way:

$n \in v(B)(\omega)$ iff $B \in \lambda(\omega(n))$

**Definition 9** *Let $M$ be a Probabilistic Structure with $Z$ it sets of atomic propositions, $\omega$ be in $Path_{ful}(M)$, $v$ be a valuation, and $\phi \in PMLO(Z)$. We define when a formula $\phi$ holds at $\omega$ in $M$ under a valuation $v$, written $M, v, \omega \models \phi$, by the following inductive clauses:*

$$
\begin{aligned}
M, v, \omega \models B(t) \quad & \text{iff} \quad v(t) \in v(B)(\omega) \\
M, v, \omega \models t < t' \quad & \text{iff} \quad v(t) < v(t') \\
M, v, \omega \models t \in X \quad & \text{iff} \quad v(t) \in v(X) \\
M, v, \omega \models \exists P_{\sim p}(\phi_1|\phi_2) \quad & \text{iff} \quad \text{if } m_1 = \mu(\{\omega' \mid \omega' \in Path^A_{ful}(M) \wedge M, v, \omega' \models (\phi_1 \wedge \phi_2)\}), \\
& \qquad \text{and } m_2 = \mu(\{\omega' \mid \omega' \in Path^A_{ful}(M) \wedge M, v, \omega' \models \phi_2\}) \\
& \qquad \text{then } m_1 \sim p \cdot m_2, \text{ for some adversary } A \\
M, v, \omega \models \exists t.\phi_1 \quad & \text{iff} \quad M, v[n/t], \omega \models \phi_1, \text{ for some } n \in \mathbb{N} \\
M, v, \omega \models \exists X.\phi_1 \quad & \text{iff} \quad M, v[N/X], \omega \models \phi_1, \text{ for some finite set of naturals } N \\
M, v, \omega \models \neg\phi_1 \quad & \text{iff} \quad \text{it is not the case } M, v, \omega \models \phi_1 \\
M, v, \omega \models \phi_1 \vee \phi_2 \quad & \text{iff} \quad \text{either } M, v, \omega \models \phi_1 \text{ or } M, v, \omega \models \phi_2
\end{aligned}
$$

It is classical fact that the set $\{\omega \mid \omega \in Path^A_{ful}(M) \wedge M, v, \omega \models \phi\}$ is measurable (see [12]).

### 3.2 Semi Markov Processes

**Definition 10** *A Probabilistic Structure $M = (S^M, \rho^M)$ is a* Semi Markov Process *if and only if $S^M$ is a Finite Automaton. If for each state $q$ there exists at most one label $a$ such that $M(q, a) \neq \emptyset$, then we call $M$ a* Markov Process.

From now on we consider the model checking problem on closed formulae in $PMLO(L)$, where $L$ is the set of atomic propositions of $M$. Hence, we say that $M$ satisfies $\phi \in PMLO(L)$, written $M \models \phi$, if and only if $M, v, \omega \models \phi$, for each valuation $v$ and $\omega \in Path_{ful}(M)$.

The set of parametrized formulae is defined similarly to the set $PMLO(Z)$ except that probabilistic operators $\exists P_{\sim p}$ with $p \in \mathbb{Q}$ are replaced with $\exists P_{\sim \alpha}$, where $\alpha$ is a parameter name. Let $\phi$ be a parametrized formula with parameters $\alpha_1, \ldots, \alpha_n$, and $\overline{p} = (p_1, \ldots, p_n)$ be a vector of rational in $\mathbb{Q}^n$. With $\phi_{\overline{p}}$ we denote the $PMLO(Z)$ -formula replacing in $\phi$ each parameter $\alpha_i$ with $p_i$. By abuse of terminology we say that $\phi$ belongs to $PMLO(Z)$ if each instance $\phi_{\overline{p}}$ of $\phi$ is in $PMLO(Z)$. A known result for Markov Processes that states the decidability of model checking for a rather large class of formulae, is the following (see [12]).

**Theorem 11** *Let $M$ be a Markov Process, $\epsilon > 0$ be a rational, and $\phi$ be a parametrized $PMLO(L)$-formula where each probabilistic operator is of the form $\exists P_{\sim \alpha}(\phi')$ where in $\phi'$ free variables are natural variables and no probabilistic operators appear. One can compute, for each parameter $\alpha_i$ in $\phi$ ($i = 1, \ldots, n$), a finite set of intervals $H_i$ not containing zero and with total length less than $\epsilon$ such that, if $\overline{p} \notin H_1 \times \cdots \times H_n$, then one can compute a $WMLO(L)$-formula $\phi'$ such that $M \models \phi_{\overline{p}}$ if and only if $M \models \phi'$.*

Since we are interested in Semi Markov Processes, the previous theorem cannot be used. More precisely, the previous theorem strongly depends on the determinism of Markov Processes. We give now a result for Semi Markov Process on *qualitative formulae* (i.e. formulae that have probabilistic operators $\exists P_{\sim p}(\phi_1|\phi_2)$ where $p \in \{0, 1\}$). Hence we restrict the values of probabilities appearing in probabilistic operators but we extend the formulae by considering second order free variables in the scope of probabilistic operators.

Let $M$ be a Semi Markov Process with set of atomic propositions $L = \{B_1, \ldots, B_k\}$, and $\phi$ a $WMLO(L)$-formula with free variables in $\{t_1, \ldots, t_n, X_1, \ldots, X_m\}$. If $S$ and $F$ are respectively the Finite Automaton and the subset of states of Theorem 3, then we can construct a Semi Markov Process $M(\phi)$ with symbols in $\Delta \times \{0, 1\}^{(n+m)}$ such that $M(\phi)$ is the Cartesian product of $M$ and $S$. The transition $((q_1, q_2), (a, \beta), (q'_1, q'_2))$ is a transition of $M(\phi)$ if and only if $(q_1, a, q'_1)$ is a transition of $M$, $(q_2, (b_1, \ldots, b_k, \beta), q'_2)$ is a transition of $S$ for some $\beta \in \{0, 1\}^{(n+m)}$, and $b_i = 1$ if and only if $B_i$ labels $q_1$ in $M$, for $i = 1, \ldots, k$. Moreover, we denote with $F^M$ the set of states $(q, q')$ of $M(\phi)$ such that $q' \in F$

An adversary is *Markovian* if it depends not on the past but only on the current state of the Semi Markov Process, i.e., if $\omega$ and $\omega'$ are two finite paths of length $k$ and $k'$ respectively such that $\omega(k) = \omega'(k')$ (the two paths have the same last state) then $A(\omega) = A(\omega')$.

**Lemma 12 ([18],[19])** *Let $M$ be a Semi Markov Process and $F$ be a subset of its states. If $A$ is an adversary for $M$ and $q$ is a state, then the set $\mathcal{P}_{M,F}(A,q) = \{\omega \in Path^A_{ful}(M,q) \mid inf(\omega) \cap F \neq \emptyset\}$ is measurable. Moreover, one can compute in polynomial time for each state $q$, the maximal value of $\mu(\mathcal{P}_{M,F}(A,q))$ for all adversaries $A$, as well as a Markovian adversary realizing this maximal value.*

**Proposition 13** *Let $M$ be a Semi Markov Process with set of atomic propositions $L$, and $\phi$ be a $WMLO(L)$-formula. One can compute a $WMLO(L)$-formula $\phi'$ with the same free variables as $\phi$ such that, for each infinite run $\omega$ of $M$ and each valuation $v$, it holds that $M, v, \omega \models \exists P_{>0}(\phi)$ if and only if $M, v, \omega \models \phi'$. The formula $\phi'$ is computed in polynomial time on the size of $M(\phi)$, and hence, the size of $\phi'$ is polynomial in the size of $M(\phi)$.*

**PROOF.** Observe that, for each run $\omega = q_0 \xrightarrow{a_0,\beta_0} \ldots$ in $Path_{ful}(M(\phi), q)$ such that $inf(\omega) \cap F^M \neq \emptyset$, there exists a natural $n_1$ such that, for each $n_2 > n_1$, $\beta_{n_2} = (0, \ldots, 0)$ because the free variables of $\phi$ are interpreted as finite sets. Let $M'$ be the Semi Markov Process obtained from $M(\phi)$ in the following way: for each symbol $a \in \Delta$, if there exists a transition $q \xrightarrow{a,\beta} q'$ with $\beta \neq (0, \ldots, 0)$, then remove all the transitions starting from $q$ where the label has a first component equal to $a$. Now the transitions have labels of the form $(a, (0, \ldots, 0))$. Using Lemma 12, one can compute the set $F^{>0}$ of states $q$ of $M'$ such that

$$max(\{\mu(\mathcal{P}_{M',F^M}(A,q) \mid A \text{ is an adversary of } M'\}) > 0.$$

Notice that $q \in F^{>0}$ iff there exists a Markovian adversary $A$ for $M'$ such that $\mu(\mathcal{P}_{M',F^M}(A,q)) > 0$. Let $\mathcal{X} = \{t_1, \ldots, t_n, X_1, \ldots, X_m\}$ be the free variables in $\phi$, and $v$ be a valuation for these variables. Clearly, $M, v, \omega \models \exists P_{>0}(\phi)$ iff there is in $M(\phi)$ a finite run with some length $n_1$ from the initial state to a state of $F^{>0}$ such that, for $i = 1, \ldots, n$, $h \leq n_1$, and $j = 1, \ldots, m$, we have that $v(t_i) = h$ iff $b_i^h = 1$, and $h \in v(X_j)$ iff $b_{n+j}^h = 1$ (($b_1^h, \ldots, b_{n+m}^h$) is the label of the $h$th transition). Hence, it is easy to construct a finite automaton $S(\phi)$ representing exactly the set of valuations $v$ of $\mathcal{X}$ such that $M, v, \omega \models \exists P_{>0}(\phi)$.

Let $S(\phi)$ be the Finite Automaton with symbols in $\{0,1\}^{(n+m)}$, with the same states as $M(\phi)$ and such that $(q, \beta, q')$ is a transition of $S(\phi)$ if and only if either $q = q' \in F^{>0}$ and $\beta = (0, \ldots, 0)$, or $q \notin F^{>0}$ and $(q, (a, \beta), q')$ is a transition of $M(\phi)$. Now the pair $(S(\phi), F^{>0})$ represents a set of valuations for the free variables of $\phi$ since $S(\phi)$ has labels in $\{0,1\}^{(n+m)}$ and each run starting from a state in $F^{>0}$ has labels of the form $(0, \ldots, 0)$.

Let $\phi'$ be the formula computed following Theorem 3 such that $\mathcal{L}(\phi') = \mathcal{L}(S(\phi), F^{>0})$. We prove that $M, v, \omega \models \exists P_{>0}(\phi)$ if and only if $M, v, \omega \models \phi'$. If it holds, then, by Lemma 12 and Theorem 3, the thesis holds.

Since for each $\omega$ and $\omega'$ it holds that $M, v, \omega \models \exists P_{>0}(\phi)$ iff $M, v, \omega' \models \exists P_{>0}(\phi)$, the satisfiability of $\exists P_{>0}(\phi)$ depends only on $M$ and $v$. Moreover, the finite Automaton $S(\phi)$ once entered in a state in $F^{>0}$ loops in this state. Actually, by definition of $M(\phi)$, by Theorem 3 and by Lemma 12, it is sufficient to enter in a state of $F^{>0}$ to have a word that describes a valuation satisfying $\exists P_{>0}(\phi)$. Hence, $\exists P_{>0}(\phi)$ is satisfied by exactly the valuation $v$ such that there exists an infinite word $a_0 a_1 \dots$ in $\mathcal{L}(S(\phi), F^{>0})$, where $a_i = (b_1^i, \dots, b_{n+m}^i)$, and such that $v(t_i) = j$ iff $b_i^j = 1$, and $v(X_i) = \{j \mid b_{i+n}^j = 1\}$. This implies that $M, v, \omega \models \exists P_{>0}(\phi)$ iff $M, v, \omega \models \phi'$.

The problem to compute $F^{>0}$ is polynomial in $M(\phi)$ (see Lemma 12). The formula $\phi'$ is computed in polynomial time in the size of $M(\phi)$ (see [17]), and hence, has polynomial size in the size of $M(\phi)$. □

The proof of $\exists P_{=1}(\phi)$ is similar to the previous one. But, since we must guarantee a probability equal to 1, we must consider subsets of states of $M(\phi)$. This because, for the previous case, to guarantee a probability greater than zero, it is sufficient that there exists a path reaching a state in $F^{>0}$. Here, since the probability must be equal to 1, we must guarantee that there exists an adversary whose all paths lead to states in $F^{=1}$.

**Proposition 14** *Let $M$ be a Semi Markov Process with set of atomic propositions $L$, and $\phi$ be a $WMLO(L)$-formula. One can compute a formula $\phi' \in WMLO(L)$ with the same free variables as $\phi$ and such that, for each infinite run $\omega$ and valuation $v$, it holds that $M, v, \omega \models \exists P_{=1}(\phi)$ if and only if $M, v, \omega \models \phi'$. The formula $\phi'$ is computed in exponential time on the size of $M(\phi)$, and the size of $\phi'$ is exponential on the size of $M(\phi)$.*

**PROOF.** In the same way as in the proof of Proposition 13, one can compute, using Lemma 12, the set $F^{=1}$ of states $q$ of $M'$ such that

$$max(\{\mu(\mathcal{P}_{M', F^M}(A, q) \mid A \text{ is an adversary of } M'\}) = 1.$$

Notice that (Lemma 12) $q \in F^{=1}$ iff there exists a Markovian adversary $A$ such that $\mu(\mathcal{P}_{M', F^M}(A, q)) = 1$. We define the Finite Automaton $S(\phi)$ where:

- The set of symbols is $\{0, 1\}^{(n+m)}$.
- States of $S(\phi)$ are subsets of states of $M(\phi)$;
- The initial state is the set containing only the initial state of $M(\phi)$;

- The transition $(G, \beta, G')$ is a transition of $S(\phi)$ if and only if either $G = G' \subseteq F^{=1}$ and $\beta = (0, \ldots, 0)$, or there exists a function $f : G \to \Delta$ such that $G'$ is the set

$$\{q' \mid \text{there exists } q \in G \text{ s.t. } (q, (f(q), \beta), q') \text{ is a transition of } M(\phi)\}.$$

The idea is that a run $\omega$ of $S(\phi)$ represents the fact that there exists an adversary $A$ and a valuation $v$ such that the possible states reachable at step $i$ are $\omega(i)$, and the infinite word that labels $\omega$ represents the valuation $v$. The function $f$ represents the choice of an adversary at a certain step. The finite Automaton $S(\phi)$, once entered in a subset of states in $F^{=1}$, loops since, by Lemma 12 and by definition of $F^{=1}$, it is sufficient to enter in a subset of $F^{=1}$ to have a word that describes a valuation satisfying $\exists P_{=1}(\phi)$. Hence $\phi'$ is the formula such that $\mathcal{L}(\phi') = \mathcal{L}(S(\phi), 2^{F^{=1}})$. The proof is similar to that of Proposition 13 by using Lemma 12 and since, if $(G, \beta, G')$ is a transition of $S(\phi)$, then there exists an adversary such that the probability to reach state $G'$ from $G$ is equal to 1.

The time to compute $F^{=1}$ is polynomial in $M(\phi)$ (see Lemma 12). The formula $\phi'$ is computed in exponential time in the size of $M(\phi)$, and has an exponential size in the size of $M(\phi)$. □

Since $\exists P_{=1}(\phi)$ is equivalent to $\exists P_{=0}(\neg\phi)$, one can replace in Proposition 13 $\exists P_{>0}(\phi)$ with $\exists P_{\sim 0}(\phi)$. Moreover, since $\exists P_{\sim 0}(\phi_1|\phi_2)$ is equivalent to $\exists P_{\sim 0}(\phi_1 \wedge \phi_2)$, by applying repeatedly Propositions 13 and 14, we have the following Theorem.

**Theorem 15** *Let $M$ be a Semi Markov Process with atomic propositions in $L$ and $\phi \in PMLO(L)$ be a qualitative formula. One can compute a formula $\phi' \in WMLO(L)$ such that $M \models \phi$ if and only if $M \models \phi'$. Hence the model checking problem for Semi Markov Processes of qualitative formulae in $PMLO(L)$ is decidable.*

## 4 Probabilistic Timed Automata

In this section, we define the class of *Probabilistic Timed Automata* and the Probabilistic Logic $PMLO(L_C)$ for Probabilistic Timed Automata.

### 4.1 Probabilistic Timed Automata

We assume a set $C$ of variables, called *clocks*. A *clock valuation* $\xi$ for a set of clocks $C$ is a function that assigns a non-negative real value to each clock. For a clock valuation $\xi$ and a time value $\tau$, $\xi + \tau$ denotes the clock valuation such that $(\xi + \tau)(x) = \xi(x) + \tau$, for any $x \in C$. Moreover, for a given set of clocks $C' \subseteq C$, with $\xi[C']$ we denote the clock assignment which sets each clock in $C'$ to 0; more precisely, $\xi[C'](x) = 0$ if $x \in C'$, and $\xi[C'](x) = \xi(x)$ otherwise.

The most general set of *clock constraints* over a set of clocks $C$, denoted $\Psi(C)$, is defined by the following grammar, where $\psi_1, \psi_2$ range over $\Psi(C)$, $x, y \in C$, $c \in \mathbb{Z}$ and $\sim \in \{<, \leq, =, \neq, >, \geq\}$.

$$\psi ::= x \sim c \,|\, x - y \sim c \,|\, \psi_1 \wedge \psi_2 \,|\, \psi_1 \vee \psi_2 \,|\, true$$

**Definition 16** *A tuple $T = (C, S, trap, \gamma)$ is a* Probabilistic Timed Automaton *if the following requirements are satisfied:*

- *$C$ is a finite set of clocks.*
- *$S = (\Delta, L, Q, q_0, Tr, \lambda)$ is a Finite Automaton. We will write $T(q, a)$ to denote the set of transitions $S(q, a)$.*
- *trap is a trap state such that $trap \notin Q$, hence, no transitions in $Tr$ argue on trap.*
- *$\gamma : Tr \to \Psi(C) \times 2^C \times (0, 1]$ is a probability condition function. If $\gamma(e) = (\psi, C', p)$, then with $cond(e)$, $res(e)$ and $prob(e)$ we denote $\psi$, $C'$ and $p$, respectively. Moreover, we require that, for all states $q \in Q$ and symbols $a \in \Delta$, we have that $\sum_{e \in T(q,a)} prob(e) = 1$.*

The trap state is entered when in a certain situation, a transition is not enabled. In this case, the probability of reaching *trap* is equal to the sum of the probabilities of the transitions non enabled. This is necessary to ensure that from $T$ we can derive a probabilistic structure. Actually, the definition of probabilistic structures requires that the sum of the probabilities of the steps enabled in a certain state w.r.t. a certain symbol is equal to 1.

Moreover, we note that the previous definition is equivalent to that given in [6]. In [6] a transition is of the form $(q, a, \psi, C', p, q')$. The case in which between $q$ and $q'$, with the symbol $a$, we have $k > 1$ transitions can be modelled with our formalism by replicating the state $q'$ (each new state can be target only with transitions with a certain fixed pair $(\psi, C')$). We note that this transformation has a polynomial cost. In a similar way we can simulate the definition of [5].

A *configuration* of $T$ is a pair $(q, \xi)$ where $q \in Q \cup trap$ and $\xi$ is a clock valu-ation on $C$. With $\mathcal{C}(T)$, we denote the set of configurations of $T$.
The *initial configuration* $s_0$ is the configuration $(q_0, \xi_0)$, where for each clock $x$ it holds that $\xi_0(x) = 0$.

We consider now the Probabilistic Structure defined by $T$.

**Definition 17** *The Probabilistic Timed Automaton $T$ defines the Probabilistic Structure $(S^T, \rho^T)$ such that the labelled transition system $S^T$ is equal to the tuple $((\mathbb{R}^{\geq 0} \times \Delta), L, \mathcal{C}(T), (q_0, \xi_0), Tr^T, \lambda^T)$ where $\lambda^T(q, \xi) = \lambda(q)$, and $((q, \xi), (\tau, a), (q', \xi'))$ is in $Tr^T$ if and only if one of the following requirements holds:*

(1) $e = (q, a, q')$ is in $Tr$, $(\xi + \tau) \models cond(e)$ and $\xi' = (\xi + \tau)[res(e)]$.
(2) $\xi' = \xi + \tau$, $q' = trap$, and there exists a transition $e = (q, a, q'') \in Tr$, for some $q''$, such that $(\xi + \tau) \not\models cond(e)$.
(3) $\xi' = \xi + \tau$ and $q = q' = trap$.

*Moreover, the probabilistic function $\rho^T$ is such that if $e = ((q, \xi), (\tau, a), (q', \xi'))$ is a transition in $Tr^T$, then*

$$\rho^T(e) = \begin{cases} prob((q, a, q')) & \text{if } q, q' \neq trap \\ \sum_{e' \in T(q,a) \text{ s.t. } \xi + \tau \not\models cond(e')} prob(e') & \text{if } q \neq trap \text{ and } q' = trap \\ 1 & \text{if } q = q' = trap \end{cases}$$

Hence, if there exists a transition $e$ with label $a$ such that $cond(e)$ holds at time $\tau$, then there exists a step at time $\tau$ with label $(\tau, a)$. The new values of clocks are incremented by time $\tau$, and the clocks in $res(e)$ are reset to 0. Moreover, there exists a step that reaches the trap state at time $\tau$ if there exists a transition $e$ with target $q''$ such that $cond(e)$ does not hold at time $\tau$. Finally, once entered in a trap state we must loop on it.

As done before, we will not make distinction between $T$ and the Probabilistic structures $(S^T, \rho^T)$ and the labelled transition system $S^T$ that $T$ defines. Hence, as an example, we will write $T \models \phi$ to denote that $(S^T, \rho^T) \models \phi$ and $Path_{ful}(T)$ to denote the set $Path_{ful}(S^T)$.

We define now the set of of predicates $L_C$.

**Definition 18** *Let $T$ be a Probabilistic Timed Automaton with a set of atomic propositions $L$ and a set of clocks $C$. We define the (infinite) set of predicate* symbols $L_C$ as follows:

$$L_C = L \cup \{\sim_c^{x,y}, \sim_c^x, \sim_c^{x,+}\}$$

where $x, y \in C$, $c \in \mathbb{N}$ and $\sim \in \{<, \leq, =, \neq, \geq, >\}$. From now on we suppose that each valuation $v$ gives to predicate symbols in $L_C$ the following interpretation:

- The set of atomic propositions $L$ has the interpretation given in Section 2.
- $(\omega, i) \in v(\sim_c^{x,y})$ if and only if $\xi_i(x) \sim \xi_i(y) + c$ where $\omega(i) = (q_i, \xi_i)$, namely the values of $x$ and $y$ in step $i$ are related by $\sim$ and $c$. The case $\sim_c^x$ requires that $\xi_i(x) \sim c$.
- $(\omega, i) \in v(\sim_c^{x,+})$ if and only if $\xi_i(x) + \tau_i \sim c$ where $\omega = (q_0, \xi_0) \xrightarrow{(a_0, \tau_0)} \ldots$, namely the value of $x$ just before the reset of step $i$ is related with $c$ by $\sim$.

Let $t$ be a natural variable and $x, y$ be two clocks, we will write $x_t \sim y_t + c$, $x_t \sim c$ and $x_t^+ \sim c$ for the formulae $\sim_c^{x,y}(t)$, $\sim_c^x(t)$ and $\sim_c^{x,+}(t)$, respectively. We note that we have not defined a predicate of the form $\sim_c^{x,y,+}$ since $x_t^+ \sim y_t^+ + c$ holds if and only if $x_t \sim y_t + c$ holds.

As for Semi Markov Processes, from now on we consider the model checking problem $T \models \phi$ such that $\phi$ is a closed formula in $PMLO(L_C)$. The decidability result on Semi Markov Processes cannot be directly used for Timed Probabilistic Automata since the Probabilistic Structure defined by a Probabilistic Timed Automaton has an infinite set of states and symbols. The fact that we have not considered relations between the values of variables in different steps is because a formula of the form $x_{t_1} \sim y_{t_2} + c$ makes the model checking problem undecidable (even if we consider formulae without probabilistic operators). Actually, let $L_{diff}$ be the set of predicates symbols $x_{t_1} \sim y_{t_2} + c$ with the semantics $(\omega, i_1, i_2) \in v(x_{t_1} \sim y_{t_2} + c)$ if and only if $\xi_{i_1} \sim \xi_{i_2} + c$, where $\omega(i_1) = (q_{i_1}, \xi_{i_1})$ and $\omega(i_2) = (q_{i_1}, \xi_{i_2})$. The following theorem states that the model checking problem for formulae in $WMLO(L_C \cup L_{diff})$ (hence formulae without probabilistic operators) is undecidable.

**Theorem 19** *It is undecidable to check whether $T \models \phi$ for a given Probabilistic Timed Automaton $T$ and a formula $\phi \in WMLO(L_C \cup L_{diff})$.*

**PROOF.** We translate the reachability problem of a 2-counter machine (that is undecidable) into the problem of checking whether $T \models \phi$.

A 2-counter machine consists of two counters $J$ and $K$, and a sequence of $n$ instructions. Each instruction may increment or decrement one of the counters, or jump, conditionally upon one of the counters being zero. After the execution of a non jump instruction, it proceeds to the next instruction. A configuration is a triple $(b, m_1, m_2)$ where $b \in [0, n-1]$ is the index of the actual instruction, $m_1$ is the value of $J$ and $m_2$ is the value of $K$. Sequences of configurations are defined in an obvious way. The problem to check whether there exists a finite sequence of configurations starting from $(0,0,0)$ such that the last configuration is equal to a given configuration $(b, m_1, m_2)$ is undecidable.

We consider the Probabilistic Timed Automaton $T$ with set of symbols $\{a\}$, set of clocks $C = \{x, pc, K, J\}$, set of states $\{q_1, q_{pc}, q_K, q_J\}$, labelling $\lambda$, such that $\lambda(q) = q$, for any state $q$, set of transitions $Tr$, and probabilistic condition function $\gamma$ such that:

- $e = (q_1, a, q) \in Tr$, for any state $q$, and $\gamma(e) = (x = 0, C, \frac{1}{4})$;
- $e = (q_y, a, q_1) \in Tr$, with $y \in C \setminus \{x\}$, and $\gamma(e) = (x = 0 \wedge pc < n, \emptyset, \frac{1}{4})$;
- $(q_y, a, q_{y'}) \in Tr$, with $y, y' \in C \setminus \{x\}$, and $\gamma(e) = (x = 1, \{x\}, \frac{1}{4})$.

The finite runs $\omega$ of $T$ are such that if $\{i_1, \ldots, i_l\}$ are the indexes of $\omega$ such that $\omega(i_j) = (q_1, \xi)$, then the triple $(\xi(pc), \xi(J), \xi(K))$ represents the configuration of the 2-counter machine at step $j$. In fact, the clock $x$ permits the clocks to assume only natural values since it is reset in each step and each condition requires that $x$ is either equal to 0 or to 1. Hence in state $q_y$ the clock $y \in \{pc, K, J\}$ is reset. In state $q_1$ we are able to read the configuration created in states $\{q_{pc}, q_K, q_J\}$.

Now we define a formula $\phi(b, m_1, m_2)$ such that $T \not\models \phi(b, m_1, m_2)$ if and only if the configuration $(b, m_1, m_2)$ is not reachable by the 2-counter machine.

Firstly we model the instructions. We show a modelling of the increment of counter $J$. The other instructions can be modelled similarly. If we are on step $t$, then the formula $\phi(t)$ equal to

$$\exists t' > t. q_1(t') \wedge pc_{t'} = pc_t + 1 \wedge J_{t'} = J_t + 1 \wedge K_{t'} = K_t \wedge \forall t'' \in (t, t'). \neg q_1(t'')$$

models the fact that in the next step w.r.t. $t$ (represented by $t'$) the counter $J$ is increased with 1. Hence, the set of sequences of length $\bar{t}$ of a 2-counter machine is modelled by the formula

$$\phi_{prog}(\bar{t}) = \forall t. \left( (t < \bar{t} \wedge q_1(t)) \Rightarrow \left( \bigvee_{i \in [1,n]} ((pc_t = i) \Rightarrow \phi_i(t)) \right) \right)$$

where $\phi_i(t)$ represents the formula which models the performing of the $i^{th}$ instruction of the 2-counter machine at step $t$.

Hence the closed formula modelling that the configuration $(b, m_1, m_2)$ is not reachable by the 2-counter machine is the following:

$$\phi(b, m_1, m_2) = \forall \bar{t}. (q_1(\bar{t}) \wedge \phi_{prog}(\bar{t})) \Rightarrow \neg (pc_{\bar{t}} = b \wedge J_{\bar{t}} = n \wedge K_{\bar{t}} = m)$$

Is is obvious that $T \not\models \phi(b, m_1, m_2)$ if and only of the 2-counter machine reaches the configuration $(b, m_1, m_2)$.

□

### 4.2 Region graph

Let us recall the notion of *region graph*. Since we consider diagonal constraints, we must consider the definition of regions given in [20] that is an extension of that given in [1].

Let $C$ be a set of clocks and $c_M$ be a natural constant. Let us consider the equivalence relation $\approx$ over clock valuations and constant $c_M$ that contains each pair of clock valuations $\xi$ and $\xi'$ such that:

- for each clock $x$, either $\lfloor \xi(x) \rfloor = \lfloor \xi'(x) \rfloor$, or both $\xi(x)$ and $\xi'(x)$ are greater than $c_M$ ($\lfloor z \rfloor$ indicates the integer part of $z$).
- for each clock $x, y$, either $\lfloor \xi(x) - \xi(y) \rfloor = \lfloor \xi'(x) - \xi'(y) \rfloor$, or both $\xi(x) - \xi(y)$ and $\xi'(x) - \xi'(y)$ fall out of the interval $[-c_M, c_M]$.
- for each pair of clocks $x$ and $y$ with $\xi(x) \leq c_M$ and $\xi(y) \leq c_M$, $fract(\xi(x)) < fract(\xi(y))$ if and only if $fract(\xi'(x)) < fract(\xi'(y))$ ($fract(z)$ indicates the fractional part of $z$).

Note that for each pair of valuations $\xi$ and $\xi'$, and for each clock constraint $\phi$ with constants enclosed in $[-c_M, c_M]$, it holds that:

$$\text{if } \xi \approx \xi' \text{ then } (\xi \models \phi \text{ iff } \xi' \models \phi).$$

A *clock region* is an equivalence class of clock valuations induced by $\approx$. We denote by $[\xi]$ the equivalence class of $\approx$ containing $\xi$. Note that the set of clock regions is finite.

A *region of* $T$ is a tuple $(q, [\xi])$ where $q$ is a state of $T$ and $\xi$ is a clock valuation on clocks $C$ of $T$. The idea is that $(q, [\xi])$ represents the set of configurations $(q, \xi')$ such that $\xi' \in [\xi]$.

### 4.3 Extended region graph with classical semantics

The classical definition of *region* used in [1] and [20] is a pair composed by a state and a clock region. Since in our logic we consider predicates in $L_C$, we must extend the classical definition of region graph. In fact the definition of $\approx$ region graph considers regions as states. Since we must distinguish the elapsing of time from the performing of a transition in such a way to valuate $x_t$ and $x_t^+$, we consider a notion of *extended region graph*. In the definition of the extended region graph, states are also marked with either a mark that represents the elapsing of time (label *time*) or a mark that represents the instantaneous performing of a transition (label *trans*).

**Definition 20** *Let $T = (C, (\Delta, L, Q, q_0, Tr, \lambda), trap, \gamma)$ be a Probabilistic Timed Automaton and $\phi$ be a $PMLO(L_C)$-formula. If $c_M$ is the smallest natural constant greater than each constant appearing in $T$ and $\phi$, and $G$ is the set of predicates in $L_C \setminus L$ appearing in $\phi$, then the* extended region graph *for $T$ and $\phi$, denoted with $R(T, \phi)$, is the Semi Markov Process $((\Delta^R, L^R, Q^R, q_0^R, Tr^R, \lambda^R), \rho^R)$ where:*

- $\Delta^R$ *is the set of symbols $\Delta \cup \{\lambda_{[\xi]} \mid [\xi]$ is a clock region w.r.t. the constant $c_M\}$.*
- $L^R$ *is the set $L \cup G \cup \{time\}$. Obviously, the atomic proposition time represents the fact that we are in a state marked by "time".*
- $Q^R$ *is the set of tuples $(q, [\xi], time)$ and $(q, [\xi], trans)$, where $(q, [\xi])$ is a region of $T$ for $c_M$. The marks time and trans represent the fact that configurations expressed by the region $(q, [\xi])$ are reached with the elapsing of time and with an instantaneous transition, respectively.*
- $q_0^R = (q_0, [\xi_0], time)$ *where $(q_0, \xi_0)$ is the initial configuration.*
- *The set of transitions $Tr^R$ is as follows:*
  · *$((q, [\xi], time), \lambda_{[\xi']}, (q', [\xi'], trans))$ is in $Tr^R$ if and only if $q = q'$ and there exist a time $\tau$ such that $[\xi'] = [\xi + \tau]$. Hence the label of the transition represents the clock region reachable with the time elapsing from the clock region represented in the source state.*
  · *$((q, [\xi], trans), a, (q', [\xi'], time))$ is in $Tr^R$ if and only if $e = (q, a, q')$ is in $Tr$ where $[\xi'] = [(\xi[res(e)])]$ and $\xi \models cond(e)$. Hence we express the configurations reachable by an instantaneous transition from a configuration expressed by $(q, [\xi])$.*
  · *$((q, [\xi], trans), a, (trap, [\xi'], time))$ is in $Tr^R$ if and only if $[\xi'] = [\xi]$ and there exists $e = (q, a, q') \in Tr$, for some $q'$, such that $\xi \not\models cond(e)$.*
  · *$((trap, [\xi], trans), a, (trap, [\xi], time))$ is in $Tr^R$, for any symbol $a$ and clock region $[\xi]$.*
- $\lambda^R$ *is such that $\lambda^R((q, [\xi], time))$ is the set $\lambda(q) \cup \{time\} \cup G'$ such that $G' \subseteq G$ and $B \in G'$ iff either $B = \sim_c^{x,y}$ and $\xi(x) \sim \xi(y) + c$, or $B = \sim_c^x$ and $\xi(x) \sim c$. Moreover, $\lambda^R((q, [\xi], trans))$ is the set $\{\sim_c^{x,+} \in G \mid \xi(x) \sim c\}$.*
- *The function $\rho^R$ is such that $\rho^R((q, [\xi], time), \lambda_{[\xi']}, (q', [\xi'], trans)) = 1$, since from each state with mark time it holds that there exists only one transition with label $\lambda_{[\xi]}$, for each clock region $[\xi]$; and if $e = ((q, [\xi], trans), a, (q', [\xi'], time))$* is a transition in $Tr^R$, then

$$\rho^R(e) = \begin{cases} prob((q,a,q')) & \text{if } q, q' \neq trap \\ \sum_{e' \in T(q,a) \text{ s.t. } \xi \not\models cond(e')} prob(e') & \text{if } q \neq trap \text{ and } q' = trap \\ 1 & \text{if } q = q' = trap \end{cases}$$

The extended region graph is a Semi Markov Process since for each $\xi_1, \xi_2 \in [\xi]$ the set of transitions enabled in $\xi_1$ is the same as those enabled in $\xi_2$, and hence is unique in the clock region $[\xi]$. This holds since for each condition $\psi$ that labels a transition it holds that if $\xi_1 \approx \xi_2$, then $\xi_1 \models \psi$ if and only if $\xi_2 \models \psi$.

We note also that a sequence of steps of an extended region graph is of the form

$$\omega' = (q_0, [\xi'_0], time) \xrightarrow{\lambda_{[\xi'_1]}} (q_0, [\xi'_1], trans) \xrightarrow{a_0} (q_1, [\xi'_2], time) \xrightarrow{\lambda_{[\xi'_3]}} \ldots,$$

namely, an alternating sequence of states with marks *time* and *trans* and hence an alternating sequence of symbols $\lambda_{[\xi]}$ and symbols in $\Delta$. Therefore, the idea is that $x_t$ must be evaluated in step $2 \cdot t$ of $R(T, \phi)$ (i.e. the $t^{th}$ state marked by *time*). Moreover, $x_t^+$, must be evaluated in step $2 \cdot t + 1$ of $R(T, \phi)$ (i.e. the $t^{th}$ state marked by *trans*).

### 4.4 Relations between Probabilistic Timed Automata and Extended Region Graph

As a consequence of results in [1], [20], [14] and [21] we have the following theorem.

**Theorem 21** *Let $T$ be a Probabilistic Timed Automaton with propositions in $L$, clocks in $C$, and $\phi \in PMLO(L_Q)$. The following facts hold.*

- *Let $\omega = (q_0, \xi_0) \xrightarrow{(\tau_0, a_0)} (q_1, \xi_1) \ldots$ be in $Path_{ful}(T)$. There exists a unique run*

$$\omega' = (q_0, [\xi'_0], time) \xrightarrow{\lambda_{[\xi'_1]}} (q_0, [\xi'_1], trans) \xrightarrow{a_0} (q_1, [\xi'_2], time) \xrightarrow{\lambda_{[\xi'_3]}} \ldots$$

*of $R(T, \phi)$ such that $\xi_i \in [\xi'_{2i}]$ and $\xi_i + \tau_i \in [\xi'_{2i+1}]$, for any $i \geq 0$. We say that $\omega'$ is the representant of $\omega$, and we denote it with $[\omega]$.*

- *Let $\omega = (q_0, [\xi_0], time) \xrightarrow{\lambda_{[\xi_1]}} (q_0, [\xi_1], trans) \xrightarrow{a_0} (q_1, [\xi_2], time) \xrightarrow{\lambda_{[\xi_3]}} \ldots$ ;*

there exists a run
$$\omega' = (q_0, \xi'_0) \xrightarrow{(\tau_0, a_0)} (q_1, \xi'_1) \ldots$$
in $Path_{ful}(T)$ such that $[\omega'] = \omega$. We say that $\omega'$ is represented by $\omega$.

It is obvious that Theorem 21 holds also if one considers finite runs finishing in states marked with *time*. Moreover, if $S$ is a set of runs of $T$, then with $[S]$ we denote the set $\{[\omega] \mid \omega \in S\}$. Theorem 21 states that the representant is unique. The following Lemma states that also the represented is unique if one considers runs defined by a certain adversary.

**Lemma 22** *Let $T$ be a Probabilistic Timed Automaton, $A$ be an adversary of $T$ and $\omega_1, \omega_2$ be two runs in either $Path^A_{fin}(T)$ or $Path^A_{ful}(T)$. If $[\omega_1] = [\omega_2]$, then $\omega_1 = \omega_2$.*

**PROOF.** Let $[\omega_1] = [\omega_2]$ and $\omega_1 \neq \omega_2$ where $\omega_i = (q_0, \xi_0^i) \xrightarrow{(\tau_0^i, a_0)} \ldots (q_n, \xi_n^i) \ldots$, for $i = 1, 2$. Let $j$ be the smallest index such that $\omega_1^{(j)} \neq \omega_2^{(j)}$. Obviously $j > 0$ since $\xi_0^1 = \xi_0^2$. Thus $\omega_1^{(j-1)} = \omega_2^{(j-1)}$ and there exists $(\tau, a) = A(\omega_1^{(j)})$ such that $\omega_1^{(j-1)} \xrightarrow{(\tau,a)} (q_j, \xi_j^1)$ and $\omega_1^{(j-1)} \xrightarrow{(\tau,a)} (q_j, \xi_j^2)$. Thus $\xi_j^1 = \xi_j^2$ which contradicts $\omega_1^{(j)} \neq \omega_2^{(j)}$. Actually, $\xi_j^1 = ((\xi_{j-1}^1 + \tau)[C_1])$ and $\xi_j^2 = ((\xi_{j-1}^2 + \tau)[C_2])$. Now $\xi_{j-1}^1 = \xi_{j-1}^2$ (since $j$ is the minimum). Therefore, let $\xi = \xi_{j-1}^1 + \tau$; we have that $\xi_j^1 = \xi[C_1] \neq \xi[C_2] = \xi_j^2$, but $\xi[C_1] \neq \xi[C_2]$ implies $[(\xi[C_1])] \neq [(\xi[C_2])]$ and hence $[\xi_j^1] \neq [\xi_j^2]$. But, this contradicts the fact that $[\omega_1] = [\omega_2]$. □

We now prove an important result concerning the probabilities between runs and represented runs.

**Lemma 23** *Let $T$ be a Probabilistic Timed Automaton, $A$ be an adversary of $T$, and $A'$ be an adversary of $R(T, \phi)$ such that $Path^{A'}_{ful}(R(T, \phi))$ is equal to $[Path^A_{ful}(T)]$. For each measurable set $S \subseteq Path^A_{ful}(T)$, it holds that $\mu(S) = \mu([S])$.*

**PROOF.** By Theorem 21 we have that for each $\omega \in Path^A_{ful}(T)$, there exists a unique $\omega' \in Path^{A'}_{ful}(R(T, \phi))$ such that $\omega' = [\omega]$. Moreover, by Lemma 22, we have that for each $\omega \in Path^{A'}_{ful}(R(T, \phi))$, there exists a unique $\omega' \in Path^A_{ful}(T)$ such that $\omega = [\omega']$. Hence there exists a bijective function between $Path^A_{ful}(T)$ and $Path^{A'}_{ful}(R(T, \phi))$. Therefore the thesis holds if, for each $\omega \in Path^A_{fin}(T)$, it holds that $\overline{\mu}(\omega) = \overline{\mu}([\omega])$, where, if $\omega = (q_0, \xi_0) \xrightarrow{(\tau_0, a_0)} \ldots (q_n, \xi_n)$, then $[\omega] = (q_0, [\xi_0], time) \xrightarrow{\lambda_{[\xi_0 + \tau_0]}} (q_0, [\xi_0 + \tau_0], trans) \xrightarrow{a} \ldots (q_n, [\xi_n], time)$.

We prove this by induction on the length of $\omega$. If $length(\omega) = 0$, then the thesis holds since $\overline{\mu}(\omega) = 1 = \overline{\mu}([\omega])$. If $length(\omega) = k > 0$, then, by definition of $R(T, \phi)$, we have that $length([\omega]) = 2 \cdot k$. Let $\omega' = \omega^{(k-1)}$; we have that $\overline{\mu}(\omega' \xrightarrow{(\tau,a)} (q', \xi'))$ is equal to $\overline{\mu}(\omega') \cdot \rho^T(((q, \xi), (\tau, a), (q', \xi')))$, where $(q, \xi)$ is the last configuration in $\omega'$. Moreover, if $e_1 = ((q, [\xi], time), \lambda_{[\xi+\tau]}, (q, [\xi + \tau], trans))$ and $e_2 = ((q, [\xi + \tau], trans), a, (q', [\xi'], time))$, then

$$\overline{\mu}([\omega]) = \overline{\mu}([\omega'] \xrightarrow{\lambda_{[\xi+\tau]}} (q, [\xi+\tau], trans) \xrightarrow{a} (q', [\xi'], time)) = \overline{\mu}(\omega') \cdot \rho^R(e_1) \cdot \rho^R(e_2).$$

Therefore the thesis holds since, by definition of $(S^T, \rho^T)$ and $R(T, \phi)$, $\rho^R(e_1) = 1$ and $\rho^T((q, \xi), (\tau, a), (q, \xi')) = \rho^R(e_2)$, and, by induction, $\overline{\mu}(\omega') = \overline{\mu}([\omega'])$. □

If $t$ is a natural variable, then with $\overline{t}$ we consider a new natural variable related to $t$. Let $v$ be a valuation; with $(2v)$ we denote the valuation such that

- for each natural variable $t$, if $v(t)$ is defined, then $(2v)(t) = 2 \cdot v(t)$ and $(2v)(\overline{t}) = 2 \cdot v(t) + 1$; otherwise both $(2v)(t)$ and $(2v)(\overline{t})$ are undefined.
- for each predicate variable $X$, if $v(X)$ is defined, then $(2v)(X) = \{2 \cdot n \mid n \in v(X)\}$; otherwise $(2v)(X)$ is undefined.

We want to prove that $T, v, \omega \models \phi$ if and only if $R(T, \phi), (2v), [\omega] \models Trans(\phi)$, where $Trans$ is a function that, given a $\phi \in PMLO(L_C)$, returns a formula in $PMLO(L^R)$, where $L^R$ is the set of atomic propositions of $R(T, \phi)$. Hence in figure 1 we give the table of translations of formulae $\phi$ in $PMLO(L_C)$.

We explain the main idea. The predicates $B \in \{\sim_c^{x,y}, \sim_c^x \mid x, y \in C \wedge c \in \mathbb{N}\}$ must be evaluated in the even steps $t$, when conditions $B \in \{\sim_c^{x,+} \mid x \in C \wedge c \in \mathbb{N}\}$ must be evaluated in step $t + 1 = \overline{t}$. When we have a formula $\exists t. \phi_1$ we must ensure to consider in $Trans(\exists t. \phi_1)$ the valuations that give an even value to $t$ (as required by definition of $(2v)$). Hence, since the proposition $time$ holds in $R(T, \phi)$ only in the even steps, with $time(t)$ we ensure that $t$ is even, and with $\overline{t} = t + 1$ we ensure that $\overline{t}$ is the successor of $t$ (as the definition of $(2v)$ requires). In a similar way the condition $\exists t.(X(t) \Rightarrow time(t))$ of $Trans(\exists X. \phi_1)$ ensures that in $X$ we have only even naturals. Now we prove the following result.

**Theorem 24** *It holds that $T \models \phi$ iff $R(T, \phi) \models Trans(\phi)$.*

**PROOF.** We prove by induction on the structure of $\phi$ of probabilistic operators that for each $\omega$ and valuation $v$, it holds that $T, v, \omega \models \phi$ if and only if $R(T, \phi), (2v), [\omega] \models Trans(\phi)$.

| $\phi \in PMLO(L_C)$ | $Trans(\phi) \in PMLO(L^R)$ |
|---|---|
| $B(t)$ with $B \in L \cup \{\sim_c^{x,y}, \sim_c^x \mid x,y \in C \wedge c \in \mathbb{N}\}$ | $B(t)$ |
| $B(t)$ with $B \in \{\sim_c^{x,+} \mid x \in C \wedge c \in \mathbb{N}\}$ | $B(\overline{t})$ |
| $t < t'$ | $t < t'$ |
| $t \in X$ | $t \in X$ |
| $\exists t.\phi_1$ | $\exists t, \overline{t}.(time(t) \wedge \overline{t} = t+1) \wedge (Trans(\phi_1))$ |
| $\exists X.\phi_1$ | $\exists X. [\forall t.(X(t) \Rightarrow time(t))] \wedge Trans(\phi_1)$ |
| $\exists P_{\sim p}(\phi_1 \mid \phi_2)$ | $\exists P_{\sim p}(Trans(\phi_1) \mid Trans(\phi_2))$ |
| $\neg \phi_1$ | $\neg Trans(\phi_1)$ |
| $\psi_1 \vee \phi_2$ | $Trans(\phi_1) \vee Trans(\phi_2)$ |

Fig. 1. The function $Trans$.

The case $\phi = B(t)$ with $B \in L$ is obvious since $Trans(B(t)) = B(t)$ and by construction of $R(T, \phi)$, Theorem 21 and by definition of $(2v)$ we have that $B \in \lambda(\omega(v(t)))$ if and only if $B \in \lambda(q)$ where $([\omega](2 \cdot v(t))) = (q, [\xi], time)$ if and only if $B \in \lambda^R(q)$ where $([\omega]((2v)(t))) = (q, [\xi], time)$. The case $\phi = B(t)$ with either $B \in \{\sim_c^{x,y}, \sim_c^x, \sim_c^{x,+} \mid x, y \in C \wedge c \in \mathbb{N}\}$ can be proved similarly. In fact the formulae in $\{\sim_c^{x,y}, \sim_c^x \mid x, y \in C \wedge c \in \mathbb{N}\}$ must be evaluated in the even steps, and hence, by definition of $(2v)$, in the step $t$. Moreover, the formulae in $\{\sim_c^{x,+} \mid x \in C \wedge c \in \mathbb{N}\}$ must be evaluated in the step $t+1$, since the value of $x_t^+$ is that expressed in the states with mark $trans$. Therefore, by definition of $(2v)$, in the step $\overline{t}$.

For the case $\phi = t < t'$, we have that $v(t) < v(t')$ iff $2 \cdot v(t) < 2 \cdot v(t')$, but by definition of $(2v)$ we have $(2v)(t) < (2v)(t')$.

The case $\phi = t \in X$ is obvious since $Trans(t \in X) = t \in X$ and by definition of $(2v)$ we have that $v(t) \in v(X)$ if and only if $2 \cdot v(t) \in \{2 \cdot n \mid n \in v(X)\}$ if and only if $(2v)(t) \in (2v)(X)$.

The case $\phi = \exists t.\phi_1$ holds by induction. Actually, $Trans(\exists t.\phi_1)$ is equal to $\exists t, \overline{t}.(time(t) \wedge \overline{t} = t+1) \wedge (Trans(\phi_1))$. Now, for some $n$, we have that $T, v', \omega \models \phi_1$ where $v' = v[n/x]$. But there exists $v''$ equal to $(2(v'))$, such that $R(T, \phi), v'', [\omega] \models time(t) \wedge \overline{t} = t+1$ and, by induction, $R(T, \phi), v'', [\omega] \models Trans(\phi_1)$. Therefore $R(T, \phi), v'', [\omega] \models Trans(\exists t.\phi_1)$. The vice versa and the case $\exists X.\phi$ can be proved similarly.

The cases $\neg\phi_1$ and $\phi_1 \vee \phi_2$ hold by induction.

We prove now the case $\exists P_{\sim p}(\phi_1|\phi_2)$.

First of all we note that $Trans(\exists P_{\sim p}(\phi_1|\phi_2)) = \exists P_{\sim p}(Trans(\phi_1)|Trans(\phi_2))$. Therefore, by induction, we have that for each $\omega$ and valuation $v$, it holds that $T, v, \omega \models \phi_i$ if and only if $R(T, \phi), (2v), [\omega] \models Trans(\phi_i)$, for $i = 1, 2$.

We prove the two implications:

- $T, v, \omega \models \exists P_{\sim p}(\phi_1|\phi_2)$ implies $R(T, \phi), (2v), [\omega] \models Trans(\exists P_{\sim p}(\phi_1|\phi_2))$

    Since $T, v, \omega \models \exists P_{\sim p}(\phi_1|\phi_2)$, we have that there exists an adversary $A$ of $T$ such that $\mu(S_1^A) \sim p \cdot \mu(S_2^A)$ where $S_1^A$ and $S_2^A$ denote the sets of runs defined by adversary $A$ and satisfying $\phi_1$ and $\phi_2$, respectively. More precisely, for $i = 1, 2$,
    
    $$S_i^A = \{\omega' \mid \omega' \in Path_{ful}^A(T) \wedge T, v, \omega' \models \phi_i\}.$$
    
    We can construct an adversary $A'$ for $R(T, \phi)$ such that for each $\omega = (q_0, \xi_0) \xrightarrow{a_0, \tau_0} \ldots (q_n, \xi_n)$ in the set $Path_{fin}^A(T)$, such that $A(\omega) = (a, \tau)$ it holds that $A'([\omega]) = \lambda_{[\xi_n+\tau]}$ and $A'(\omega') = a$, where
    
    $$\omega' = [\omega] \xrightarrow{\lambda_{[\xi_n+\tau]}} (q_n, [\xi_n + \tau], trans).$$
    
    $A'$ is defined in any way for $\omega \in Path_{fin}(R(T, \phi))$ such that $\omega$ represents no run in $Path_{fin}^A(T)$.
    
    This construction is possible thanks to Lemma 22 that states that there exists at most one run for each representant.
    
    Therefore, by Theorem 21 it holds that $\omega \in Path_{ful}^A(T)$ if and only if $[\omega] \in Path_{ful}^{A'}(R(T, \phi))$, and hence $Path_{ful}^{A'}(R(T, \phi)) = [Path_{ful}^A(T)]$.
    
    Now, by induction, we have that $S_1^{A'}$ is equal to $[S_1^A]$ and $S_2^{A'}$ is equal to $[S_2^A]$, where
    
    $$S_i^{A'} = \{\omega \mid \omega \in Path_{ful}^{A'}(R(T, \phi)) \wedge R(T, \phi), (2v), \omega \models Trans(\phi_i)\},$$
    
    for $i = 1, 2$.
    
    Hence to have the thesis it is sufficient to prove that $\mu(S_1^A) = \mu(S_1^{A'})$ and $\mu(S_2^A) = \mu(S_2^{A'})$, but this holds by Lemma 23. Therefore we have proved that if there exists an adversary $A$ of $T$ such that $\mu(S_1^A) \sim p \cdot \mu(S_2^A)$, then there exists an adversary $A'$ of $R(T, \phi)$ such that $\mu(S_1^{A'}) \sim p \cdot \mu(S_2^{A'})$.

- $R(T, \phi), (2v), [\omega] \models Trans(\exists P_{\sim p}(\phi_1|\phi_2))$ implies $T, v, \omega \models \exists P_{\sim p}(\phi_1|\phi_2)$

    Since $R(T, \phi), (2v), [\omega] \models \exists P_{\sim p}(Trans(\phi_1)|Trans(\phi_2))$, we have that there exists an adversary $A$ of $R(T, \phi)$ such that $S_1^A \sim p \cdot S_2^A$ with $S_1^A$ and $S_2^A$

denoting the sets of runs defined by adversary $A$ and satisfying $Trans(\phi_1)$ and $Trans(\phi_2)$, respectively. More precisely, for $i = 1, 2$,

$$S_i^A = \{\omega' \mid \omega' \in Path_{ful}^A(R(T, \phi)) \wedge R(T, \phi), (2v), \omega' \models Trans(\phi_i)\}$$

We construct the adversary $A'$ for $T$ such that for each run $\omega = (q_0, \xi_0) \xrightarrow{a_0, \tau_0} \ldots (q_n, \xi_n)$ in $Path_{fin}(T)$ it holds that $A'(\omega) = (a, \tau)$ if $A([\omega]) = \lambda_{[\xi_n + \tau]}$ and $A(\omega') = a$ where

$$\omega' = [\omega] \xrightarrow{\lambda[\xi_n + \tau]} (q_n, [\xi_n + \tau], trans).$$

Therefore, by Theorem 21 it holds that $\omega \in Path_{ful}^{A'}(T)$ if and only if $[\omega] \in Path_{ful}^A(R(T, \phi))$, and hence $Path_{ful}^A(R(T, \phi)) = [Path_{ful}^{A'}(T)]$. Now, by induction, we have that $S_1^A$ is equal to $[S_1^{A'}]$ and $S_2^A$ is equal to $[S_2^{A'}]$, where

$$S_i^{A'} = \{\omega \mid \omega \in Path_{ful}^{A'}(T) \wedge T, v, \omega \models \phi_i\},$$

for $i = 1, 2$.

Hence to have the thesis it is sufficient to prove that $\mu(S_1^A) = \mu(S_1^{A'})$ and $\mu(S_2^A) = \mu(S_2^{A'})$, but this holds by Lemma 23.

Therefore we have proved that if there exists an adversary $A$ of $R(T, \phi)$ such that $\mu(S_1^A) \sim p \cdot \mu(S_2^A)$, then there exists an adversary $A'$ of $T$ such that $\mu(S_1^{A'}) \sim p \cdot \mu(S_2^{A'})$.

□

Hence by Theorems 24 and 15 we have the following results.

**Corollary 25** *Let $T$ be a Probabilistic Timed Automaton with propositions in $L$ and clocks in $C$. If $\phi \in PMLO(L_C)$ is a qualitative formula, then one can compute a formula $\phi' \in WMLO(L)$ such that $T \models \phi$ if and only if $R(T, \phi) \models \phi'$.*

**Theorem 26** *Let $T$ be a Probabilistic Timed Automaton with propositions in $L$ and clocks in $C$. If $\phi \in PMLO(L_C)$ is a qualitative formula, then it is decidable whether $T$ satisfies $\phi$.*

### 4.5 Urgent semantics

In this section, we can consider also an urgent semantics where transitions must be taken as soon as possible.

**Definition 27 (urgent semantics)** *The Probabilistic Timed Automaton $T= (C, (\Delta, L, Q, q_0, Tr, \lambda), trap, \gamma)$ defines with urgent semantics the Probabilistic Structure $(S^U, \rho^U)$ such that the labelled transition system $S^U$ is equal to the tuple $((\mathbb{R}^{\geq 0} \times \Delta), L, \mathcal{C}(T), (q_0, v_0), Tr^U, \lambda^U)$ where $((q, \xi), (\tau, a), (q', \xi'))$ is in $Tr^U$ if and only if one of the following requirements holds:*

(1) *$e = (q, a, q')$ in $Tr$, $\xi + \tau \models cond(e)$ and $\xi' = (\xi + \tau)[res(e)]$. Moreover, if $e' \in T(q, a')$, then $\xi + \tau' \not\models cond(e')$, for each $\tau' < \tau$ and $a' \in \Delta$. We call these kinds of steps* real urgent transitions.

(2) *$\xi' = \xi + \tau$, $q' = trap$, and there exists a transition $e = (q, a, q'') \in T(q, a)$ such that $(\xi + \tau) \not\models cond(e)$. Moreover there exists a real urgent transition $(((q, \xi), (\tau, a), (q'', \xi'')))$ in $Tr^U$, for some $(q'', \xi'')$.*

(3) *$\tau = 0$, $\xi' = \xi$, $q' = trap$ and, for each time $\tau'$ and symbol $a'$ and configuration $(q'', \xi'')$, there exists no real urgent transition $((q, \xi), (\tau', a'), (q'', \xi''))$.*

*Moreover, $\lambda^U$ and $\rho^U$ are defined as $\lambda^T$ and $\rho^T$ of Definition 20, respectively.*

The definition of step with urgent semantics requires that a step can be performed by using a transition with the minimum possible delay. Hence we have called these kinds of steps real urgent transitions since they are performed by using a real transition in $Tr$. Moreover, a step can lead into a trap state. We have two cases. In the former case, there exists a real urgent step in the meantime, hence a trap state is reached because in the meantime some transition is not enabled. In the latter case there is no real urgent step that is performable, and hence, the step is performed with time 0.

We will write $T \models_u \phi$ to denote $(S^U, \rho^U) \models \phi$.

As a consequence of Theorem 19 we have the following corollary.

**Corollary 28** *Given a Probabilistic Timed Automaton $T$ and a formula $\phi \in WMLO(L \cup L_{diff})$, it is undecidable to check whether $T \models_u \phi$.*

**PROOF.** Since the Probabilistic Timed Automaton defined in the proof of Theorem 19 has the same set of runs with urgent semantics, then the thesis holds. □

The extended region graph $R^u(T, \phi)$ corresponding to a urgent semantics can be constructed from $R(T, \phi)$ by deleting the transitions that do not satisfy the urgent semantics. These transitions can be easily computable. First of all the set of valuations in $[\xi]$ can be written as the convex space represented by a linear formula $\pi_{[\xi]}$ on real variables $\{x_{old} \mid x \in C\}$, where $x_{old}$ represents the value of clocks $x \in C$ before the elapsing of time (see [22]). This holds also for condition $cond(e)$ labelling transition $e$. Actually, $cond(e)$ can be written as a linear formula $\pi_e$ on real variables $\{x_{new} \mid x \in C\}$, where $x_{new}$ represents the value of clocks $x \in C$ after the time elapsing. Now, for each region $(q, [\xi])$, it is sufficient to construct the linear formula

$$\exists \{x_{nex}, x_{old} \mid x \in C\}. \bigwedge_{a \in \Delta} \bigwedge_{e \in T(a,q)} x_{nex} = x_{old} + \tau \wedge \pi_{[\xi]} \wedge \pi_e,$$

where $\tau$ represents the time in which the transition can be taken. By using quantifier elimination algorithm in [23], we have an equivalent formula of the form $\tau \in I$, where $I$ a finite union of intervals. If $I$ has minimum $\tau_M$, the only reachable clock region from $[\xi]$ is $[\xi + \tau_M]$. If $I$ has not a minimum, then the only reachable state is $trap$ with a time $\tau = 0$.

Following the proof of Theorem 24, we can prove also the following theorem.

**Theorem 29** *Let $T$ be a Timed Probabilistic Automaton with propositions in $L$ and clocks in $C$ and $\phi \in PMLO(L_C)$. It holds that $T \models_u \phi$ iff $R^u(T, \phi) \models Trans(\phi)$.*

Hence by Theorems 29 and 15 we have the following results.

**Corollary 30** *Let $T$ be a Timed Probabilistic Automaton with propositions in $L$ and clocks in $C$. If $\phi \in PMLO(L_C)$ is a qualitative formula, then one can compute a formula $\phi' \in WMLO(L)$ such that $T \models_u \phi$ if and only if $R^u(T, \phi) \models \phi'$.*

**Theorem 31** *Let $T$ be a Timed Probabilistic Automaton with propositions in $L$ and clocks in $C$. If $\phi \in PMLO(L_C)$ is a qualitative formula, then it is decidable whether $T$ satisfies $\phi$ with urgent semantics*

Now, for urgent semantics one can consider cases in which the region graph $R^u(T, \phi)$ is a Markov Process.

**Lemma 32** *Let $T$ be a Probabilistic Timed Automaton. If $T$ is such that for each state $q$ there exists at most one symbol $a$ such that the set $T(q, a) \neq \emptyset$, then the extended region graph $R^u(T, \phi)$ is a Markov Process.*

**PROOF.** Let $q$ be a state of $T$ and $[\xi]$ be a clock region. For states with mark $time$ we have that the definition of urgency requires that if $[\xi']$ is the clock region reachable form $[\xi]$ with the minimum time possible and such that for $\xi'$ there exists at least one real urgent step, then there exists only one transition $e$ with source state $(q, [\xi], time)$ and $e$ is labelled with $\lambda_{[\xi']}$. For states with mark $trans$, by hypothesis, we have that for each state $q$ there exists at most one symbol $a$ such that $T(q,a) \neq \emptyset$, but this implies that for each $r = (q, [\xi], trans)$ there exists at most one symbol $a$ such that $R^u(T, \phi)(q, a) \neq \emptyset$.

□

Hence by Theorems 11 and 29, and by Lemma 32, we have the following results.

**Corollary 33** *Let $T$ be a Probabilistic Timed Automaton, $\epsilon > 0$ be a rational, and $\phi \in PMLO(L_C)$ be a parametrized formula where each probabilistic operator is of the form $\exists P_{\sim \alpha}(\phi')$ where in $\phi'$ free variables are natural variables and no probabilistic operators appear. One can compute, for each parameter $\alpha_i$ in $\phi$ ($i = 1, \ldots, n$), a finite set of intervals $H_i$ not containing zero and with total length less than $\epsilon$ such that, if $\overline{p} \notin H_1 \times \cdots \times H_n$, then one can compute a formula $\phi' \in WMLO(L)$ such that $T \models_u \phi_{\overline{p}}$ iff $R^u(T, \phi_\alpha) \models \phi'$.*

**Theorem 34** *Let $T$ be a Probabilistic Timed Automaton, $\epsilon > 0$ be a rational, and $\phi \in PMLO(L_C)$ be a parametrized formula where each probabilistic operator is of the form $\exists P_{\sim \alpha}(\phi')$ where in $\phi'$ free variables are natural variables and no probabilistic operators appear. One can compute, for each parameter $\alpha_i$ in $\phi$ ($i = 1, \ldots, n$), a finite set of intervals $H_i$ not containing zero and with total length less than $\epsilon$ such that, if $\overline{p} \notin H_1 \times \cdots \times H_n$, then it is decidable whether $T$ satisfies $\phi$ with urgent semantics.*

## 5 An example

We model a synchronous distributed system where fault tolerance is solved by replicating the service (see [2]). We suppose that a faulty entity behaves arbitrarily, hence, it can give a correct answer or a bad answer or a delayed answer. A service is replicated $n$ times. Each replica is indexed with values in $\{0, \ldots, n-1\}$. The client requires a service from one replica called *primary*. At each instant the primary is unique (initially the primary is the replica with index 0.) After received a request by a client, the primary forwards the request to each other replica (called *backup*). Each replica that receives the forwarded request, sends the answer to the client. The client considers as correct the answer that it has received by more replicas. If the client does not receive answers before a certain time-out, then it supposes that the actual primary is faulty and so it broadcasts the request to each replica. Each replica different from the primary that receives the request directly from the client, after answering to the client, elects as the new primary the replica $i + 1 \mod n$ where $i$ is the faulty primary. In such a case the faulty replicas are restart.

The set of atomic propositions is $\{finish, correct, faulty_0, \ldots, faulty_{n-1}\}$. The proposition $finish$ means that either the client has received an answer from each replica or the time out is over. The proposition $correct$ means that the client has received the correct answer. The proposition $faulty_i$ represents the fact that the replica $i^{th}$ is faulty.

The set of symbols $\Delta$ is equal to the set

$$\{req_C, req_P, req_{Prec}\} \cup \{prim_i \mid i = 0, \ldots, n-1\} \cup \{ans_i, ans_i^w, compute_i \mid i = 0, \ldots, n-1\}$$

The symbols $req_C, req_P, req_{Prec}$ represent, respectively, the request of the client to the primary, the request of the primary to the back–ups, and, the request of the client to the replicas when the time-out is passed. The symbol $prim_i$ represents that the new primary is the replica $i^{th}$. Finally, $ans_i$ and $ans_i^w$ represent, respectively, the correct and wrong answer of the replica $i^{th}$. $compute_i$ represents the fact that replica $i^{th}$ is computing the answer.

From now on, with $Pr$ we denote the set $\{prim_i \mid i = 0, \ldots, n-1\}$.

In the example we model, for clarifying system behaviors, we will use $CCS$ notation, more precisely, we will write $a$ when the symbol is "read" by the system and $\bar{a}$ when the symbol is "provided" by the system.

Beforehand, by following definition in [3], we define a notion of product between two Probabilistic Timed Automata.

**Definition 35** *Let $T_1 = (C_1, S_1, trap_1, \gamma_1)$ and $T_2 = (C_2, S_2, trap_2, \gamma_2)$ be two Probabilistic Timed Automata such that $C_1 \cap C_2 = \emptyset$, and $S_i = (\Delta_i, L_i, Q_i, q_0^i, Tr_i, \lambda_i)$, for $i = 1, 2$. The product between $T_1$ and $T_2$, denoted $T_1 \otimes T_2$, is the Probabilistic Timed Automaton $(C, (\Delta, L, Q, q_0, Tr, \lambda), trap, \gamma)$ such that*

- $C = C_1 \cup C_2$;
- $\Delta = \Delta_1 \cup \Delta_2$;
- $L = L_1 \cup L_2$;
- $Q = Q_1 \times Q_2$;
- $q_0 = (q_0^1, q_0^2)$;
- $Tr = T_0 \cup T_1 \cup T_2$ where:
  - $T_0 = \{((q_1, q_2), a, (q_1', q_2')) \mid a \in \Delta_1 \cap \Delta_2, (q_1, a, q_1') \in Tr_1, (q_2, a, q_2') \in Tr_2\}$;
  - $T_1 = \{((q_1, q_2), a, (q_1', q_2)) \mid a \in \Delta_1 \setminus \Delta_2, (q_1, a, q_1') \in Tr_1\}$;
  - $T_2 = \{((q_1, q_2), a, (q_1, q_2')) \mid a \in \Delta_2 \setminus \Delta_1, (q_2, a, q_2') \in Tr_2\}$;
- $\lambda(q_1, q_2) = \lambda(q_1) \cap \lambda(q_2)$, for any $q \in Q_1$ and $q_2 \in Q_2$;
- $\gamma(((q_1, q_2), a, (q_1', q_2'))) = P_0 \cup P_1 \cup P_2$ where
  - $P_0 = \{(\psi_1 \wedge \psi_2, C_1' \cup C_2', p_1 \cdot p_2 \mid a \in \Delta_1 \cap \Delta_2, (\psi_1, C_1', p_1) \in \gamma_1, (\psi_2, C_2', p_2) \in \gamma_2\}$;
  - $P_1 = \{(\psi, C', p \mid a \in \Delta_1 \setminus \Delta_2, (\psi, C', p) \in \gamma_1\}$;

· $P_2 = \{(\psi, C', p \mid a \in \Delta_2 \setminus \Delta_1, (\psi, C', p) \in \gamma_2\}$;

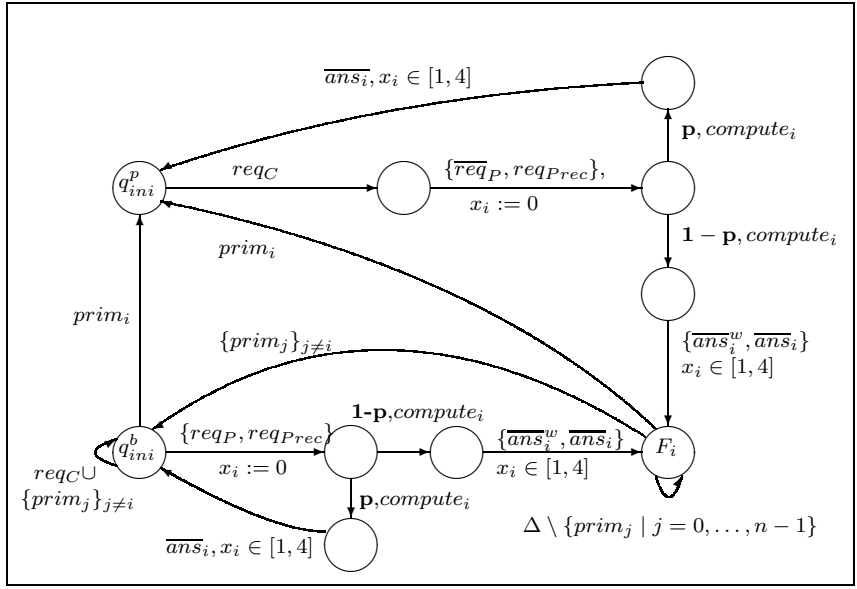

Fig. 2. The replica $i^{th}$

The replica $i^{th}$ (denoted $replica_i$) is equal to the Timed Probabilistic Automata in figure 2.

The behaviors starting from state $q_{ini}^b$ model the case in which the replica $i^{th}$ is a back–up. Now, after receiving a request from the primary ($req_P$), in a time in $[1, 4]$, it computes its answer ($\overline{ans_i}$). If it receives the request directly from the client ($req_{Prec}$), then it supposes that the primary is faulty and, after sending the answer to the client, it elects with the other non faulty replicas the new primary.

The behaviors starting from state $q_{ini}^p$ model the case in which the replica $i^{th}$ is the primary. Now, after receiving a request from the client ($req_C$), the primary sends the request of the client to each back–up ($\overline{req}_P$) and, in a time in $[1, 4]$, it computes its answer ($ans_i$).

The state $F_i$ represents the fact that the replica is faulty. The replica becomes faulty after performing a certain operation with a probability equal to $p$ (hence with probability $1-p$ the primary is not faulty after a certain operation). When the replica is faulty it can answer either correctly ($ans_i$) or incorrectly ($ans_i^w$). We suppose that the replica becomes faulty during the computation.

After the restarting, to be the back–up, the replica $i^{th}$ must read that a replica $j$, with $j \neq i$, is the primary ($\{prim_j\}_{j \neq i}$). To be the primary the replica $i^{th}$ must be initialized by the symbol $prim_i$ representing that the primary is the replica $i^{th}$.

We consider a labeling such that the proposition $faulty_i$ labels only the state

$F_i$.

The initial state of the replica 0 is $q_{ini}^p$, and, the initial state of the other replicas is $q_{ini}^b$.

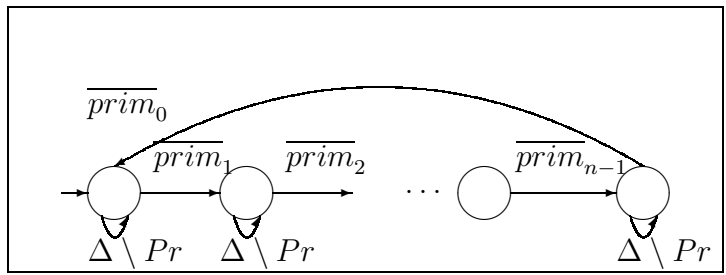

Fig. 3. The index of the primary

In figure 3 we describe the Probabilistic Timed Automaton managing the index of primary (denoted *manager*).

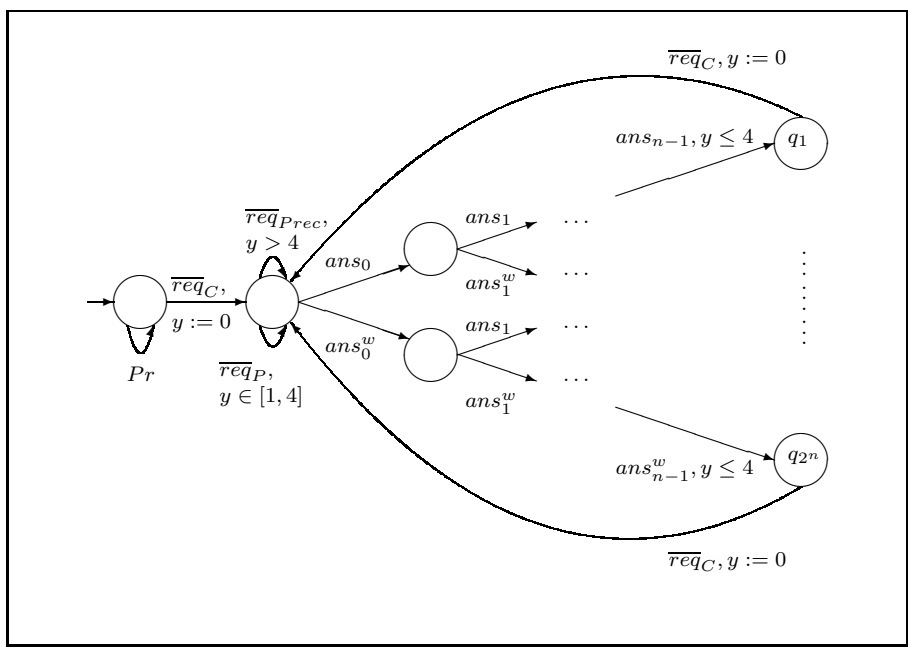

Fig. 4. The Client

In figure 4 we describe the behavior of the *client*. The client requires a service to the primary. If, after a time greater than 4, it does not receive any answer, then it supposes that the primary is faulty and sends the request to each back–up. After sending the request, the client receives an answer from each replica. The wrong answer $ans_i^w$ represents both a wrong answer and a non received answer from $replica_i$. The states $q_1, \ldots, q_{2^n}$ are the reachable states storing the answers received form each replica. We suppose the proposition *correct* labels only states $q_i$, with $1 \leq i \leq 2^n$, for which the number of correct answers needed to reach $q_i$ is greater than $\frac{n}{2}$. Moreover, the proposition $finish$ labels only states $\{q_1, \ldots, q_{2^n}\}$.

The whole system $S$ is the product of the Probabilistic Timed Automata $client$, $manager$ and $replica_0, \ldots, replica_{n-1}$.

We denote with $non\_faulty(t)$ the property

$$\bigvee_{I \subseteq [0,n-1] \text{ s.t. } |I| > \frac{n}{2}} \bigwedge_{i \in I} \neg faulty_i(t),$$

ensuring that at least $\frac{n}{2}$ replicas are not faulty. The property $\neg non\_faulty(t)$ is denoted with $faulty(t)$. Moreover, with $non\_trap$ we denote the property ensuring that the system $S$ is not in the trap state, more precisely, $non\_trap = \forall t.good(t)$ where the atomic proposition $good$ labels each state of $S$ (except the trap state).

The Probabilistic Timed Automaton $S$ enjoys the following properties:

- $\forall t. \forall P_{=1}((finish(t) \Rightarrow correct(t)) | non\_faulty(t) \wedge good(t))$, namely, if the number of faulty replicas is less than $\frac{n}{2}$, then the client receives the correct answer.
- $\forall P_{=1} \left( non\_trap \Rightarrow \left( \forall t.correct(t) \Rightarrow \bigwedge_{i \in [0,n-1]} \neg faulty_i(t) \Rightarrow ((x_i)_t \leq 4) \right) \right)$, namely the answers of the non faulty replicas are received in a time less than or equal to 4.
- $\exists P_{>0}(\exists t.faulty_0(t) \wedge \cdots \wedge faulty_{n-1}(t))$, namely there exists the possibility that each replica becomes faulty.
- $\forall P_{>0} \left( non\_trap \Rightarrow (\forall t.faulty(t) \Rightarrow \exists t' > t.non\_faulty(t)) \right)$, namely the system always has the possibility to have more than $\frac{n}{2}$ non faulty replicas (ensuring the correctness of the answer received by the client).

## 6 Discussion

In this paper we have considered the model checking problem of a logic with probabilities for Semi Markov Processes and for Probabilistic Timed Automata. The logic considered extends the Weak Monadic Second Order Logic with probabilistic operators and formulae on values of clocks in a certain step. We have proved decidability results for the class considering qualitative properties.

In this paper we have not considered two important features: repeated states and progress. We discuss now how to treat them. Now, if one is interested to consider only runs that go infinitely many times states through a certain set $F$, then it is sufficient to consider the formula $\forall t.\exists t'.t' > t \wedge rep(t')$, where the atomic proposition $rep$ labels the states in $F$.

Decidability results shown for Probabilistic Timed Automata, have consequences also in the non probabilistic case. In fact Theorem 24 implies that the model checking problem of a formula in $WMLO(L_C)$ for a non probabilistic Timed Automaton is decidable. Finally, we compare the decidable classes defined in this paper with those known in the literature. Different works deal with logics with probabilities for Markov Processes (see [24], [25] and [26], and [27] for a survey). These logics are extension of linear and branching temporal logics. In [12] it is proved that in $pCTL$ (probabilistic branching temporal logic) there is no formula equivalent to $\phi = \exists t.P_{=1}(B(t))$ (this result can be easily extended also for linear temporal logics). The formula $\phi$ means that there exists a certain step $n$ such that with probability 1 each run at step $n$ satisfies $B$. As a consequence of this fact we have that there are qualitative formulae in $PMLO(L_C)$ non expressible with the probabilistic branching time temporal logic defined in [5]. On the other hand, in [5] the authors consider also non qualitative properties. Hence the two classes are incomparable.